\documentclass{aa}

\usepackage{natbib}
\usepackage{graphicx}
\usepackage{wrapfig}
\usepackage{lscape}
\usepackage{color}
\bibpunct{(}{)}{;}{a}{}{,} 

\newcommand{\beq}{\begin{equation}}
\newcommand{\eeq}{\end{equation}}
\newcommand{\beqa}{\begin{eqnarray}}
\newcommand{\eeqa}{\end{eqnarray}}

\begin{document}

\title{Observations of a pulse driven cool polar jet 
by SDO/AIA}

\author{Abhishek Kumar Srivastava\inst{1,2} \and Kris~Murawski\inst{2}}

\institute{Aryabhatta Research Institute of Observational Sciences (ARIES), Manora Peak, Nainital-263 129, Uttarakhand, India\\
             \email{aks@aries.res.in}
           \and
            Group of Astrophysics, UMCS, ul. Radziszewskiego 10, 20-031 Lublin, Poland\\
              \email{kmur@kft.umcs.lublin.pl}}
         
\date{Received / Accepted }

\abstract
{We observe a solar jet at north polar coronal hole (NPCH) using 
SDO AIA 304 \AA \ image data on 3 August 2010. The jet rises 
obliquely above the solar limb and then retraces its
propagation path to fall back. 
}
{We numerically model this observed solar jet by implementing
a realistic (VAL-C) model of solar temperature.}
{We solve two-dimensional ideal magnetohydrodynamic equations numerically
to simulate the observed solar jet. We consider a localized velocity pulse
that is essentially parallel to the background magnetic field lines and 
initially launched at the top of
the solar photosphere. The pulse
steepens into a shock at higher altitudes, which
triggers plasma perturbations that exhibit the
observed features of the jet. The typical direction of 
the pulse also clearly exhibits the leading front of the
moving jet.
}
{Our numerical simulations reveal that a large amplitude initial velocity pulse
launched at the top of the solar photosphere produces in general
the observed properties of the jet, e.g., upward and backward average velocities, 
height, width, life-time, and ballistic nature.
}
{The close matching between the jet observations and numerical
simulations provides first strong evidence for the formation of this
jet by a single velocity pulse. The strong velocity pulse
is most likely generated by the low-atmospheric reconnection in the polar 
region which results in triggering of the jet. The downflowing material of the
jet most likely vanishes in the next upcoming velocity pulses from lower solar atmosphere, 
and therefore distinctly launched a single jet upward in the solar atmosphere is observed.
}

\keywords{Magnetohydrodynamics (MHD) -- Sun: atmosphere-- Sun : corona}

\titlerunning{SDO/AIA Polar Jet}

\authorrunning{A.K. Srivastava, K. Murawski}
\maketitle
\section{Introduction}
\begin{figure*}
\begin{center}
\includegraphics[scale=0.9, angle=90]{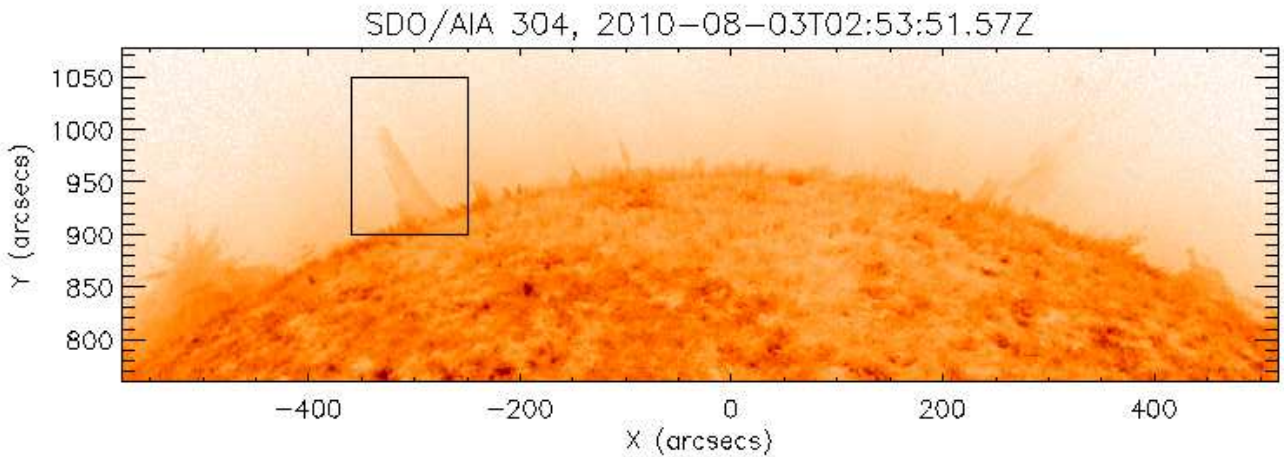}
\caption{\small
The field-of-view of the north polar coronal hole (NPCH) as observed 
by SDO/AIA 304 \AA\ on 03 August 2010 at 02:53:51 UT. The (X$_{cen}$, Y$_{cen}$)
was (-29.618$"$, 918.631$"$), while the full FOV was (1094$"$, 317$"$)
wide with a resolution of 0.6$"$ per pixel. The black box shows the
triggering of a solar jet that exhibits an exceptional phenomena of its 
most probable formation by a velocity pulse.
}
\label{fig:initial_profile}
\end{center}
\end{figure*}
Solar jets are the typical collimated and short lived transient ejecta in the solar atmosphere
that are significant in mass and energy transport at short spatio-temporal scales. Based 
on the spatial sizes, life times, and transient nature, the jets are classified 
in the form of various phenomena, e.g., anemone jets (Shibata et al. 2007), EUV jets
(Innes et al. 1997), spicules (Bohlin et al. 1975), X-ray jets (Shibata et al. 1992), and penumbral
jets (Katsukawa et al. 2007). The solar surges are cool jets typically formed with a plasma usually 
evident in H$_{\alpha}$ and other chromospheric lines, and they
are supposed to be triggered by magnetic reconnection processes (Brooks et al. 2007).
However, solar surges are mostly associated with the flaring regions
and the places of transient activities at the Sun.
Georgakilas et al. (1999) have also observed the polar surges which were the
cool surge-like jets over polar caps formed without any transient phenomenon.
However, they were having association with macro-spicules, and due to
the short-lived nature they could not attain the greater heights in off-limb
polar corona.

Other short spatio-temporal scale jets (e.g., dynamic fibrils, mottles, 
and spicules) are also significant low-atmospheric phenomena
to understand its energy and mass transport. 
Using high-resolution observations from 1 m Swedish Solar Telescope and advanced numerical
modeling, Hansteen et al. (2006) 
have shown that in the active regions these jets are a natural consequence of upwardly propagating slow-mode
magnetoacoustic shocks generated due to the leakage of p-mode sub-photospheric oscillations.
The active region fibrils are also most likely formed by chromospheric shock waves that occur 
when convective flows and global oscillations leak into the chromosphere along
the field lines of magnetic flux concentrations (De Pontieu et al. 2007a).
De Pontieu et al. (2004) have also reported for the first time the quasi-periodic rise and
fall of the spicule material due to the leakage of p-mode oscillations from 
the sub-photospheric regions.
De Pontieu et al. (2007b) have 
pointed out 
that typical type-I spicules are found to be driven by shock waves 
that form when global oscillations and convective flows
leak into the upper atmosphere along magnetic field lines on 3-7 minute timescales. They 
have also found the signature of type-II spicules, which are much
more dynamically formed within 10 s and are very thin (200 km wide) with the lifetimes of 10-150 s.
These type of spicules seem to be rapidly heated to transition region temperatures and eject material through the
chromosphere at speeds of 50-150 km s$^{-1}$ . The properties of Type II spicules suggest a formation process that
is a consequence of magnetic reconnection, typically in the vicinity of magnetic flux concentrations in plage and
network regions. 
Recently, Morton et al. (2011) have reported a unique signature of the relationship
between spicule-like structures and the polar jet formation using the SDO/AIA
observations. However, the relationships between various types of jets are still 
not well explored.

In various types of these jet phenomena, the coronal and polar 
coronal jets are very classic transient phenomena that have been observed from Skylab and
Yohkoh era in soft X-rays (Shimojo \& Shibata 2000), followed by
Solar and Heliospheric Observatory (SoHO) (Wang et al. 1998), Transition Region 
and Coronal Explorer (TRACE) (Alexander \& Fletcher 1999), and Hinode (Kamio et al. 2007
; Filippov et al. 2009). 
The magnetic reconnection is found to be one of the most likely mechanisms 
for the generation of typical coronal jets (Shibata et al. 2007), while the reconnection
generated waves may also be present with the jets during their propagation (Yokoyama \& Shibata 1995, 1996).
Pariat et al. (2009) have recently simulated the propagation of kink motion along the 
magnetic field lines around a rotating and closed magnetic field region. This dynamics
is most likely associated with the kink instability in the jet formation region, 
which may lead its upward motion in the model solar atmosphere.
The transversal waves have also been recently detected in association with the solar jet events
at small spatial scales (e.g., He et al. 2009; Kamio et al., 2010, and references cited there).
However, the evidence of Alfv\'en waves has also been detected
in the large-scale polar X-ray jets (Cirtain et al., 2007). Therefore, the magnetic reconnection
and associated waves may be the most likely drivers to energize such type
of the jets at various spatio-temporal scales. However, the exact mechanism
is not well established in the context of the very complex motion of the various types of
solar jets.

Recently, Murawski \& Zaqarashvili (2010) have reported the formation of solar spicules
by a pulse launched within the chromosphere. Their numerical simulations show
that the strong initial pulse may lead to the quasi-periodic rising of 
the chromospheric material into the lower corona in the form of spicules.
They have explained the observed speed, width, and heights of type I spicules, 
as well as observed multi-structural and bi-directional flows with a quasi-periodic
rise and fall at 3-5 min time scales due to the consecutive shocks. 
The reconnection generated velocity pulses may also drive the solar jets. However, such 
models along with their observational supports are not yet available in the context of
solar jets. Therefore, it may be 
worth to explore the launching of solar jet due to the formation of a pulse
in the lower solar atmosphere. 

In the present paper, we find a unique 
observational signature of a jet in the north polar coronal hole
which shows its dynamics as driven by a velocity pulse. We numerically simulate the 
similar kind of jet triggered by a strong amplitude velocity pulse
in the lower solar atmosphere, which in general closely mimics the observed properties of the jet.
In Sec 2, we describe the observational data and analyses. We report the numerical
model in Sec 3, and results of numerical simulation in Sec 4. 
We present the discussion and conclusions in the last section.
\begin{figure*}
\centering
\mbox{
\includegraphics[scale=0.52, angle=90]{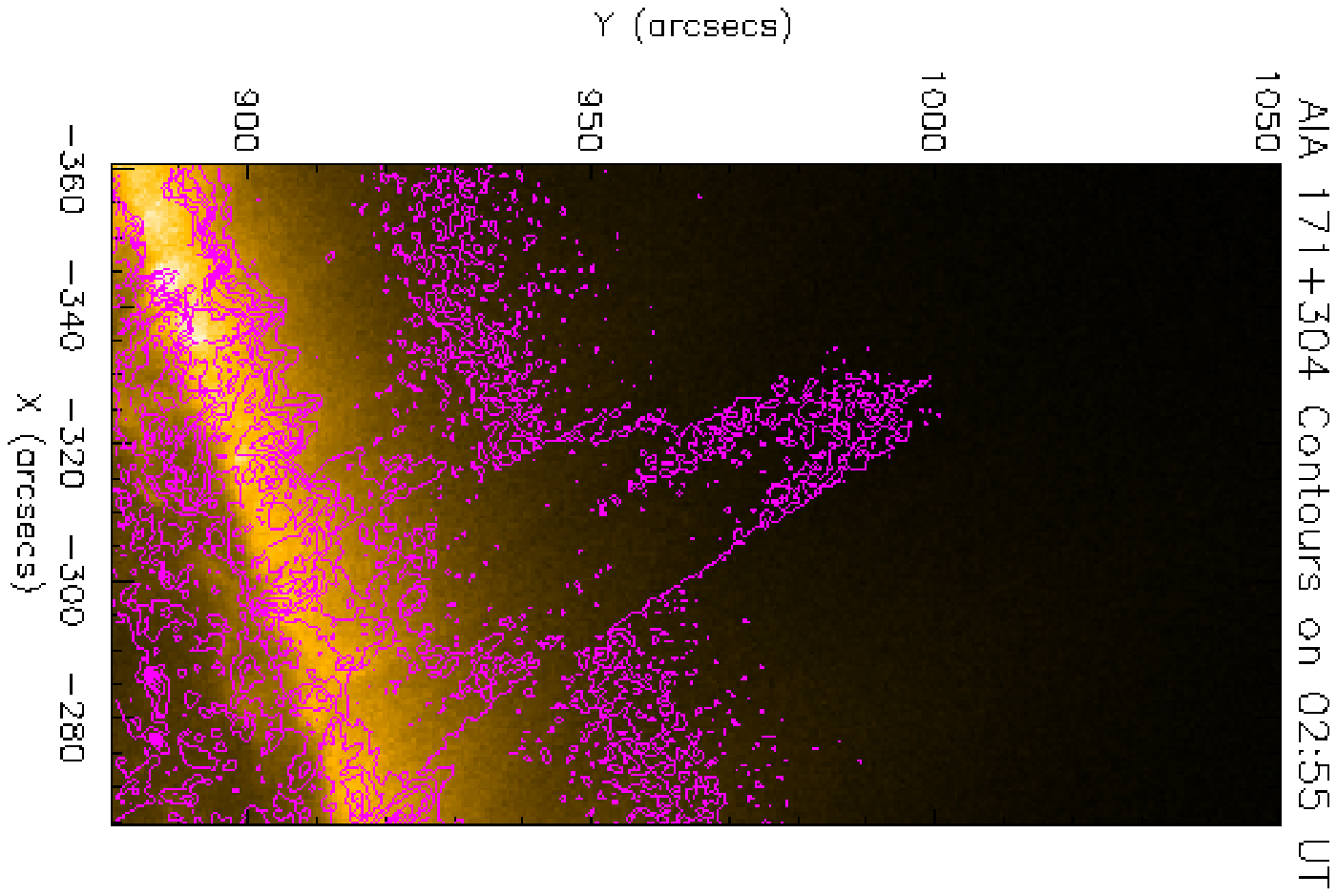}
\includegraphics[scale=0.53, angle=90]{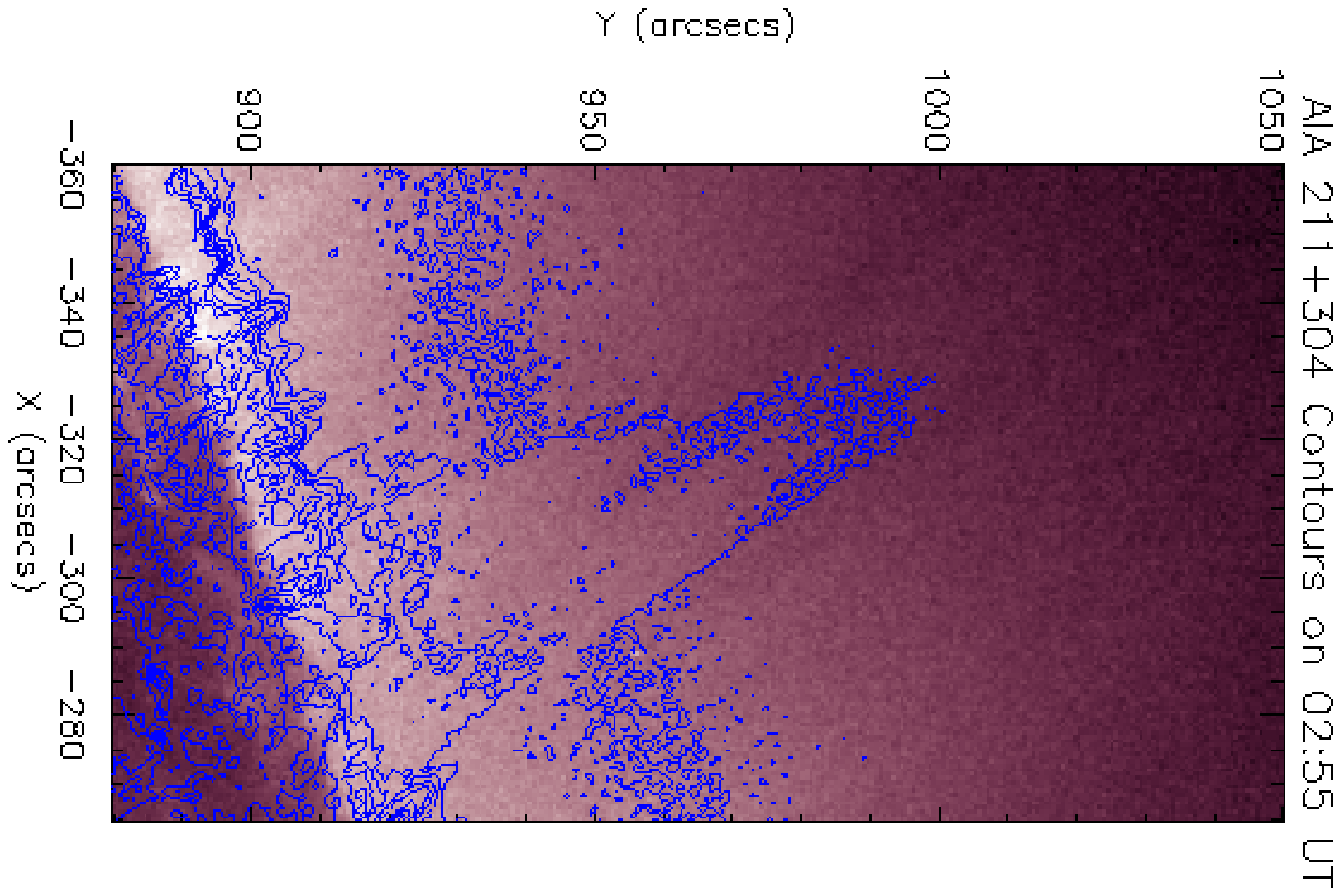}}
\caption{\small
The SDO/AIA 171 \AA\ (left) and 211 \AA\ (right) coronal images overlaid by
SDO/AIA 304 \AA\ intensity contours. It is clearly evident that the coronal filters
does not reveal any sensitivity to the jet plasma, which is observed in the
He II 304 \AA\ filter.}
\label{fig:JET-PULSE}
\end{figure*}
%
\section{Observations and results}\label{SECT:OBS}
We use a time-series data of a solar jet at north polar coronal hole 
as observed in 304 \AA \ filter of Atmospheric 
Assembly Imager (AIA) onboard the Solar Dynamics Observatory (SDO)
on 3 August 2010 during 02:41:14 UT--03:02:50 UT.
The SDO/AIA has a typical resolution of 0.6$"$ per pixel
and the highest cadence of 12 s, and it observes the
full solar disk in three UV (1600 \AA, 1700 \AA, 4500 \AA)
and seven EUV (171 \AA, 193 \AA, 211 \AA, 94 \AA, 304 \AA, 335 \AA
, 131 \AA) wavelengths (Lemen et al. 2011). Therefore, it provides 
the unprecedented observations of multi-temperature, high-resolution,
and high-temporal plasma dynamics all over the Sun.
The field-of-view of the north polar coronal hole as observed 
by SDO/AIA 304 \AA\ on 03 August 2010 at 02:53:51 UT is shown in Fig.~1.
The (X$_{cen}$, Y$_{cen}$)
was (-29.618$"$, 918.631$"$), while the full FOV was (1094$"$, 317$"$)
wide with the resolution of 0.6$"$ per pixel. The black box shows the
triggering of a solar jet off the polar limb.
The time series has been obtained by the SSW cutout service at LMSAL, USA, which
is corrected for the flat-field and spikes. We run aia\_prep subroutine 
of SSW IDL also for further calibration and cleaning of the time
series data. 

We observe a very unique jet formation in the east-north
limb of the polar coronal hole, which seems to be driven by a velocity pulse
launched in the lower solar atmosphere. The jet has been launched slightly obliquely 
from the background open field lines of the polar coronal
hole (cf., Fig. 2).
The life-time of the jet was observed as $\sim$21 min, which is the 
typical life time of the coronal jets previously observed
in the Sun and cataloged by Nistic\'o et al. (2009). We observe a typical
jet that reaches in the corona off the polar limb. However,
the jet material was only visible in He II 304 \AA\ filter, which
is sensitive to the comparatively cool temperature of 1$\times$10$^{5}$ K.
The jet material could not be evident in the typical TR/coronal filters (e.g., Fe IX 
171 \AA\ , 211 \AA\ filters) of SDO/AIA, which are sensitive to the plasma temperature $\geq$10$^{6}$ K.
Fig.~2 shows the SDO/AIA 171 \AA\ (left) and 211 \AA\ (right) coronal images overlaid by
SDO/AIA 304 \AA\ intensity contours at a time of the maximum rise
of the jet. It is clearly evident that the coronal filters
do not exhibit any sensitivity to the jet plasma, which is only observed in the
He II 304 \AA\ filter. This scenario persist during whole life-time
of the observed jet. The unique jet is a very faint event 
even captured in 304 \AA\ as most of the strong emissions come
from spicule borders at this wavelength also.
However, we may not rule out some very weak amount of coronal emissions 
also that could not be distinguished in the diffused background off the polar limb.
Therefore, we abbreviate this jet as a cool jet as its plasma mostly emits the radiations 
captured by the AIA filters that are sensitive to cooler plasma emissions,
 e.g., 304 \AA . There may also be the possibility of some traces of
emissions in other SDO/AIA filters (e.g., 1600 \AA\ ) that
are sensitive to the emissions from comparatively cooler 
plasma.
This means that the jet material was typically formed by comparatively 
cooler plasma compared to the background coronal plasma, and it moves
in form of plasma column that does not exchange any heat with the
ambient coronal plasma. This was the unique property of the observed
cool jet off-limb in the polar coronal hole. 
Our observed cool ejecta reaches at a higher height (72 Mm) with a  
life-time of 21 min and larger width of 20 Mm. We, therefore, abbreviate it as a cool polar jet,
rather as an evidence of any polar surge like ejecta.
\begin{figure*}
\centering
\mbox{
\includegraphics[scale=0.52, angle=90]{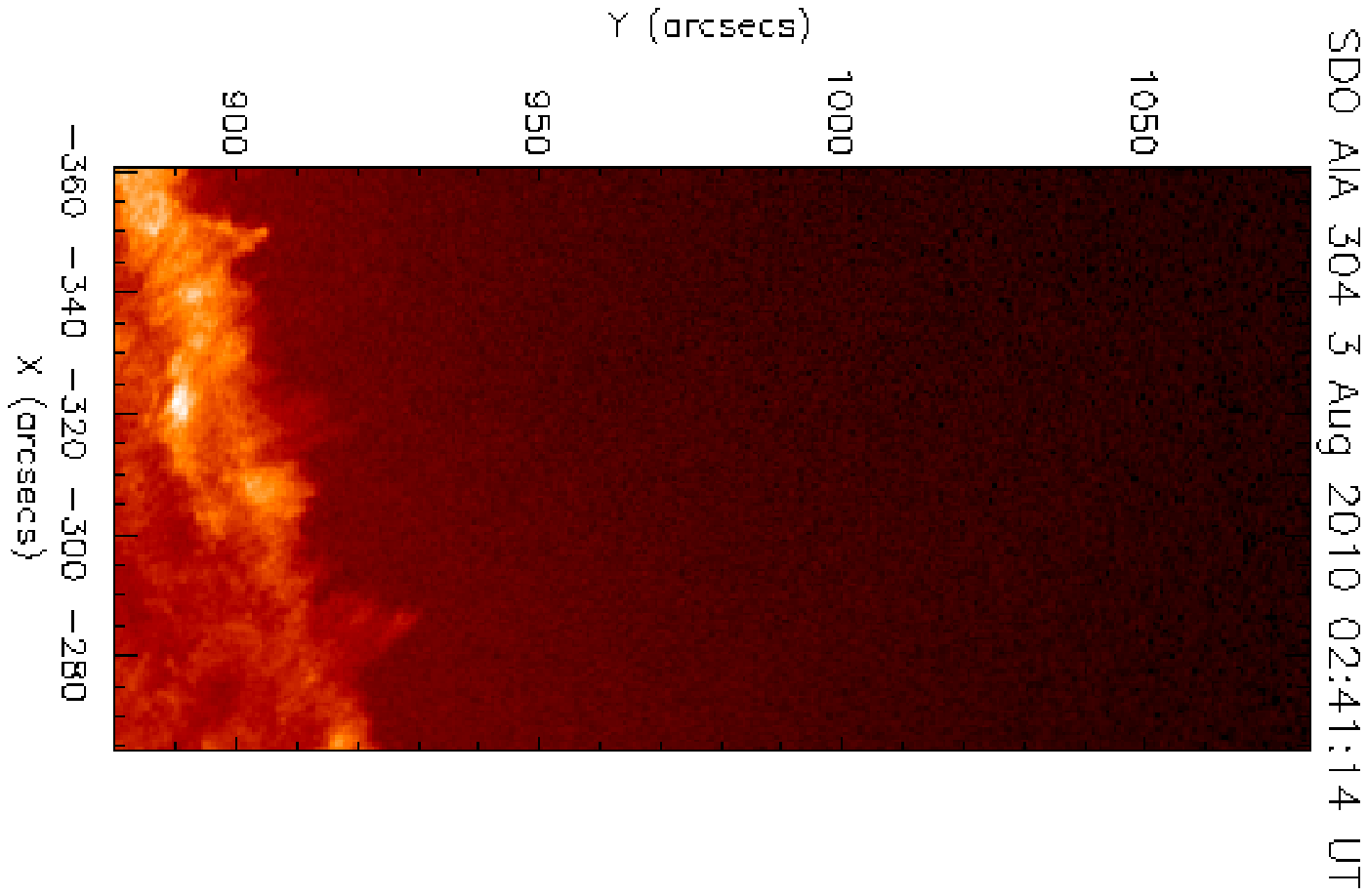}
\includegraphics[scale=0.52, angle=90]{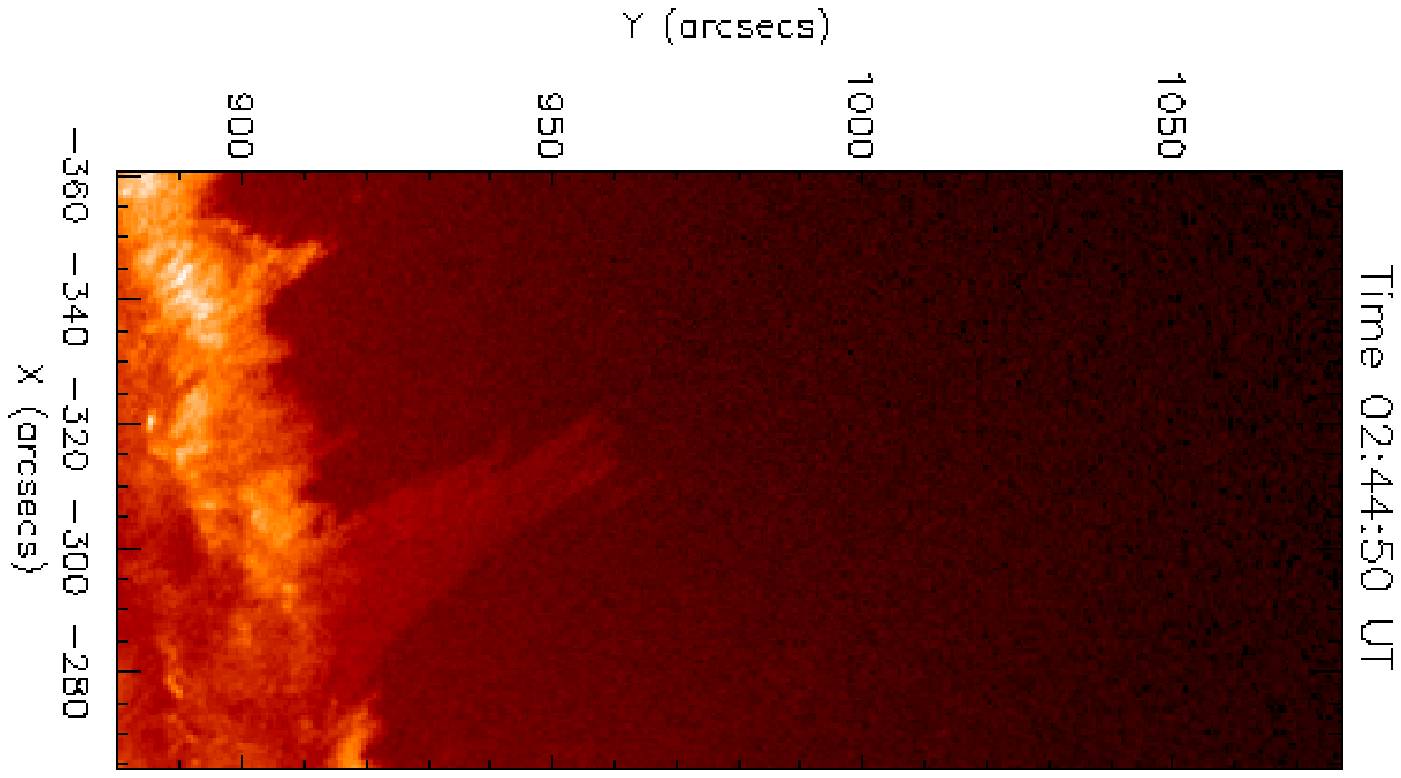}
\includegraphics[scale=0.52, angle=90]{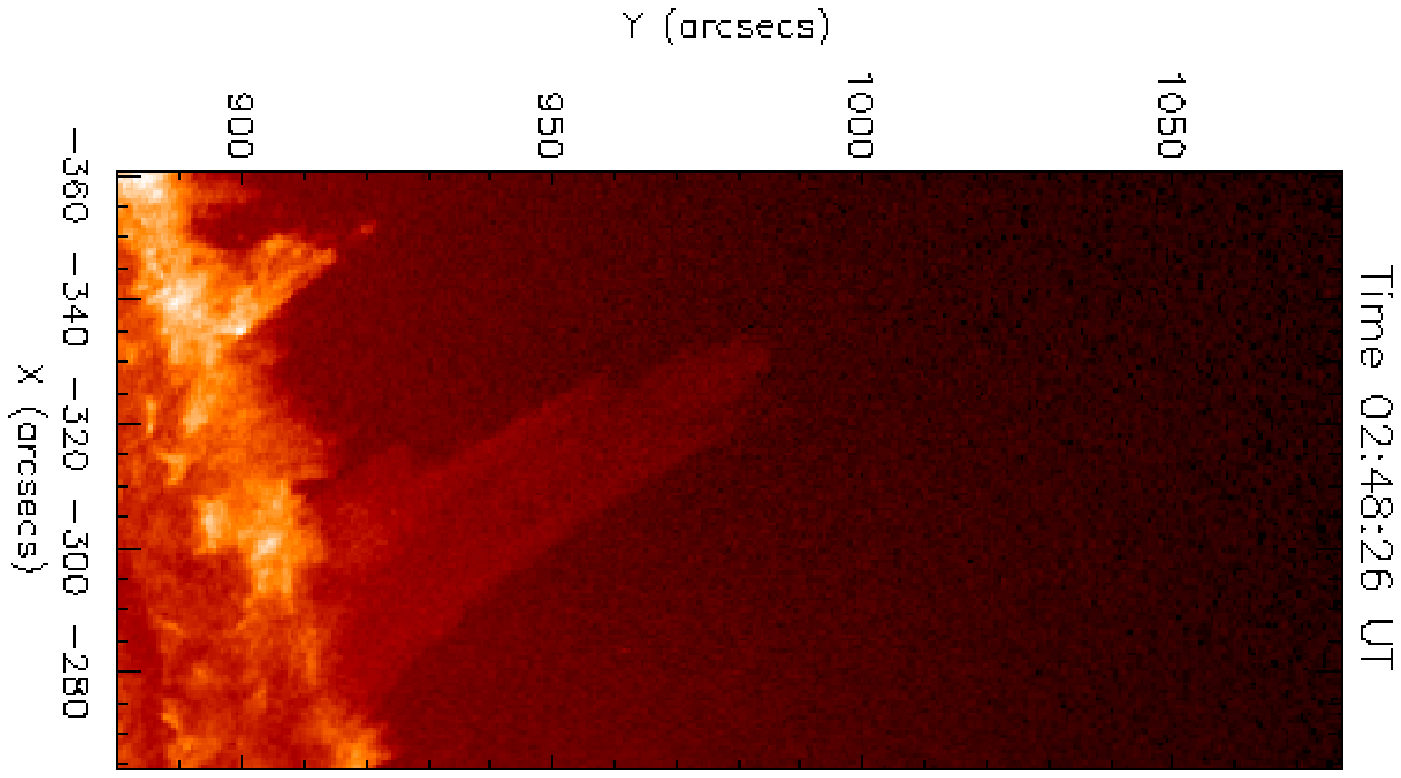}}
\mbox{
\includegraphics[scale=0.53, angle=90]{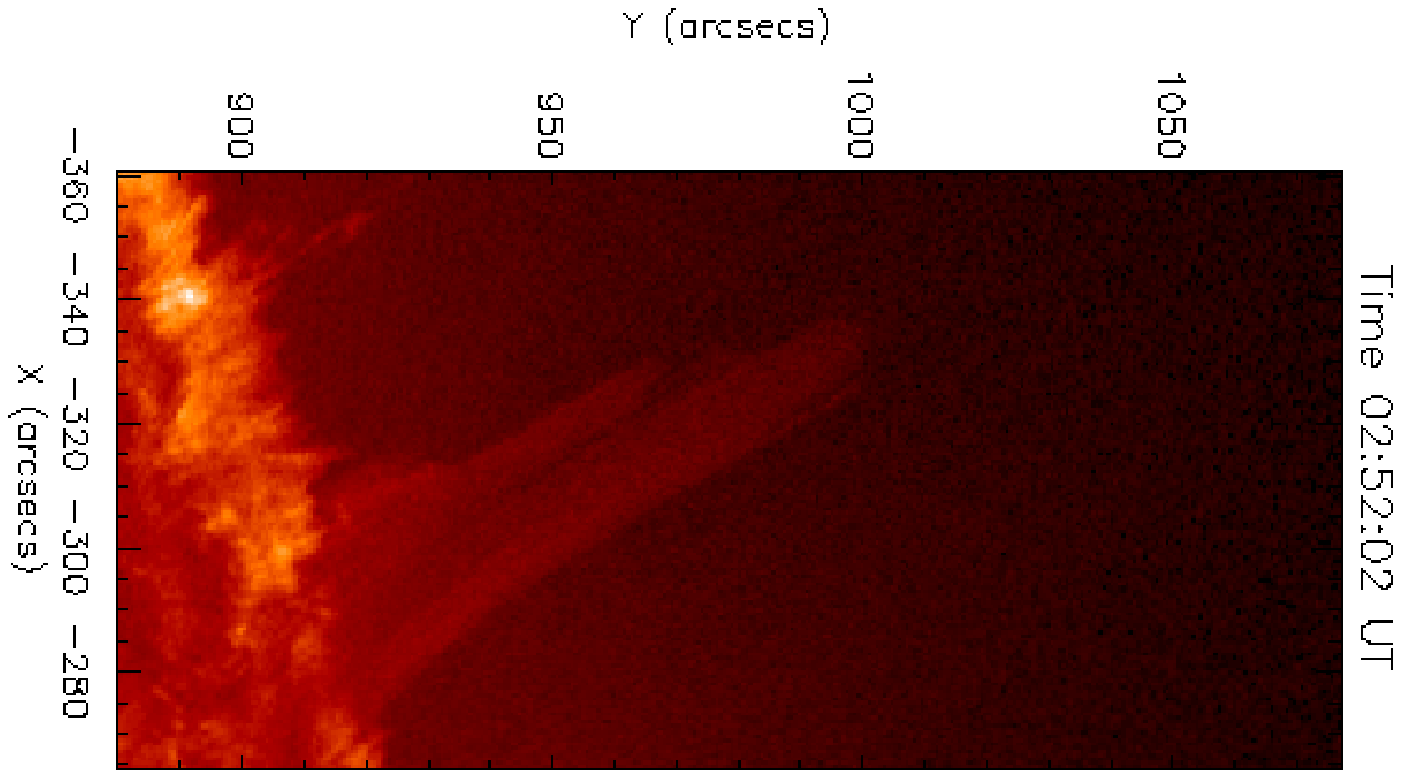}
\includegraphics[scale=0.53, angle=90]{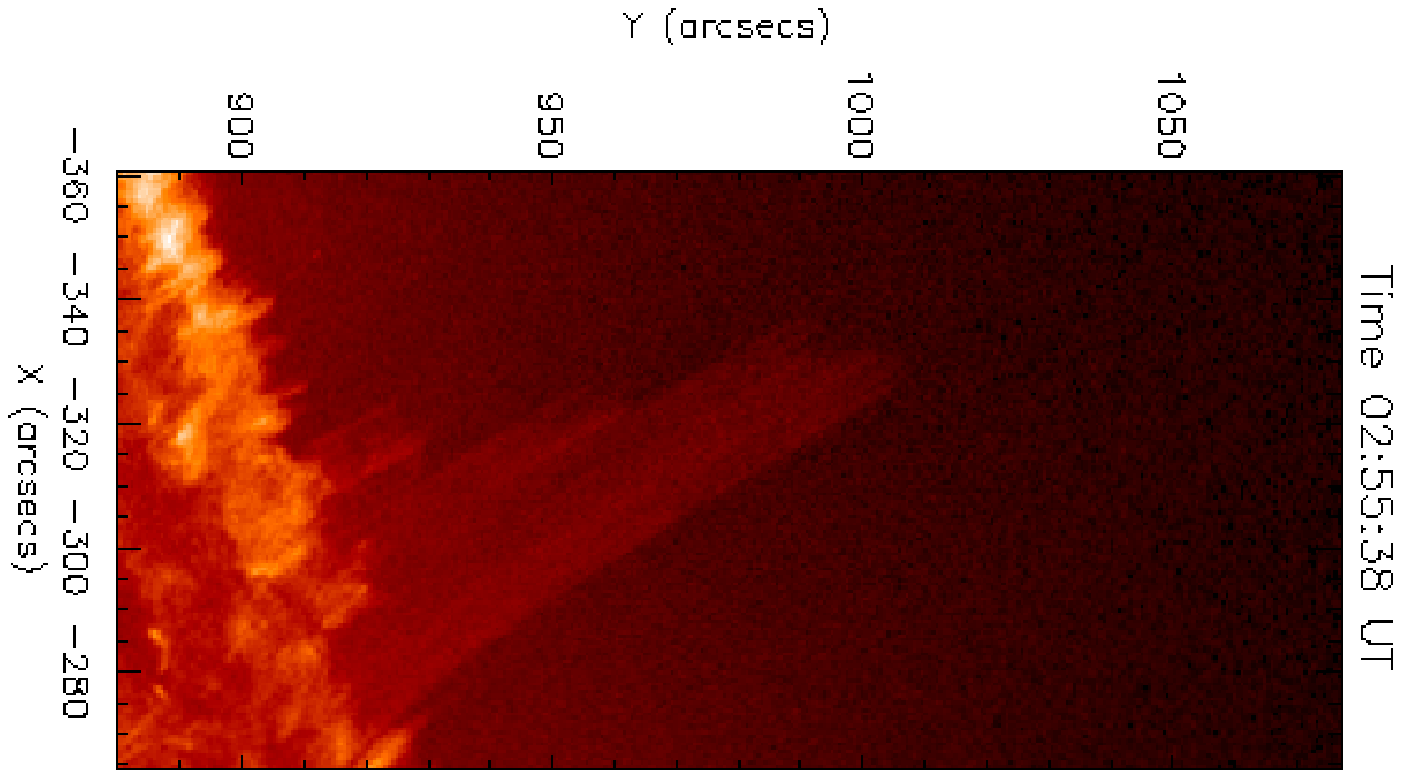}
\includegraphics[scale=0.53, angle=90]{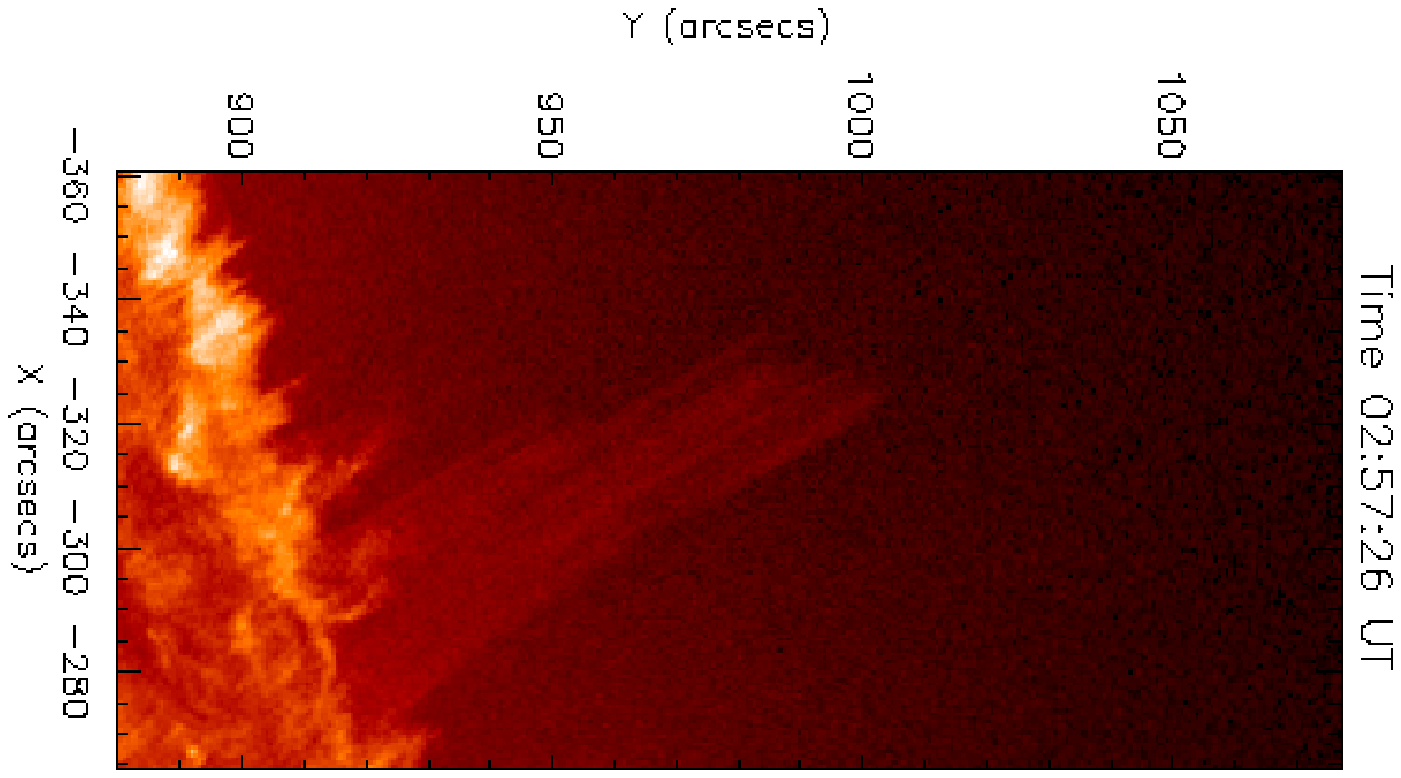}}
\mbox{
\includegraphics[scale=0.53, angle=90]{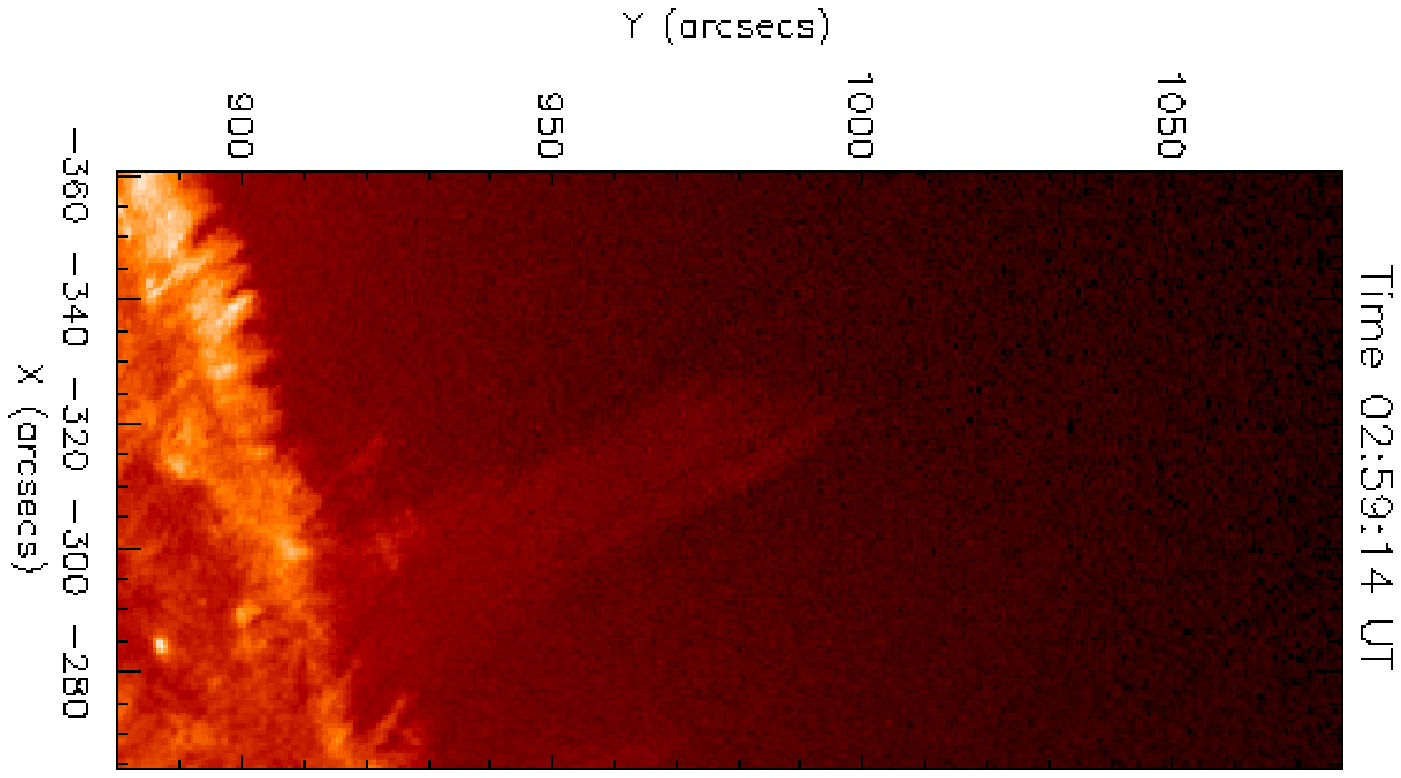}
\includegraphics[scale=0.53, angle=90]{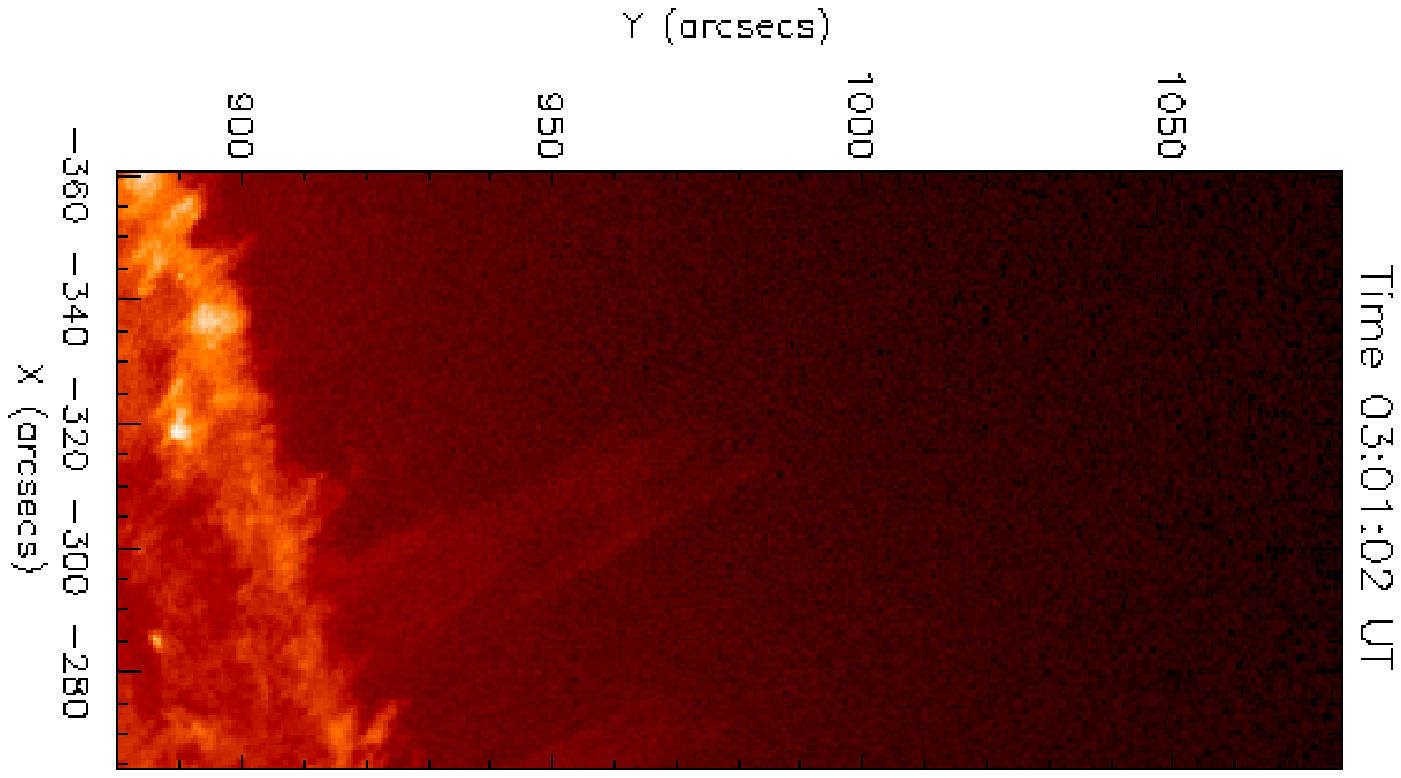}
\includegraphics[scale=0.53, angle=90]{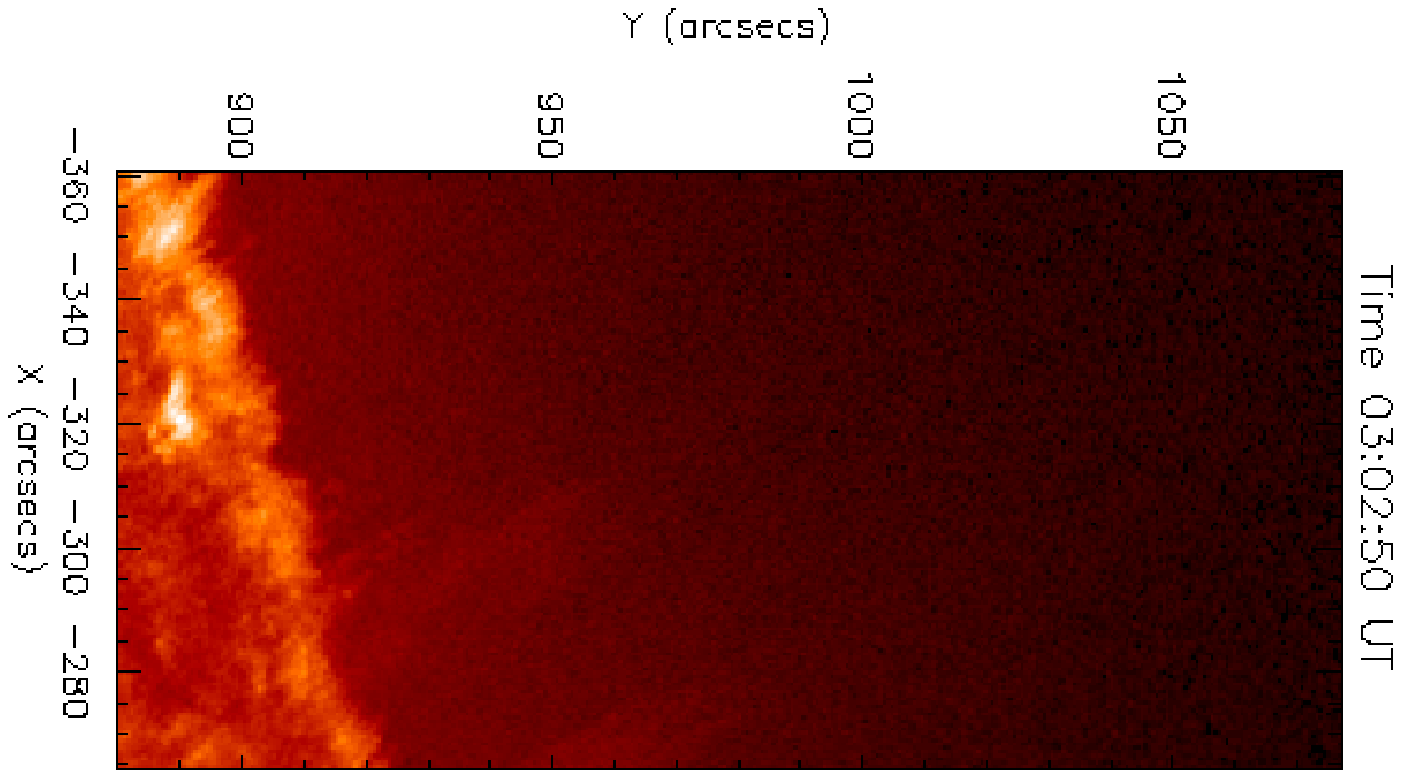}}
\caption{\small
The SDO/AIA 304 \AA\ time sequence of a cool jet propagation above the solar limb in the north-east polar coronal hole that
clearly shows the upward motion of the plasma and its downward motion along the same path.
}
\label{fig:JET-PULSE}
\end{figure*}

The jet starts moving up
above the solar limb on $\sim$02:41 UT and reaches at a maximum height
of $\sim$72 Mm at 02:52 UT. Therefore, its average rising speed is $\sim$110 km s$^{-1}$ that
is the typical speed of the jet by which it rose in the corona at a maximum 
height of $\sim$72 Mm in the first $\sim$648 s of the time (Fig. 3). Although, the width
slightly varies during the course of time at the base of the jet, 
its
shape is approximately uniform during its rise-up. Its width at the base is 
measured approximately as $\sim$20 Mm.
The unique evidence this jet presented is its rise straight 
in the corona off-limb in first 10-11 min of its life time, and then its 
fall back from the same path in rest of the duration of its life time.
Some unique observations of chromospheric jet 
by Hinode/SOT Ca II H filter revealed
fine structuring as well as transversal
motion that was associated with the helicity
injection in the upper solar atmosphere (Liu et al. 2009).
However, the spatial resolution of SDO/AIA (0.6$"$) is indeed 
lower 
compared to SOT (0.2$"$), and secondly
the observed cool jet is very fainter to clearly 
identify such fine structuring. Moreover, 
the observed intriguing chromospheric jet (Liu et al. 2009)
was associated with a small flare at 
the western limb, while in the present
case we observe the cool jet in polar coronal hole.
Therefore, these two cases are entirely different, and we do not observe such type of oscillating
threads as well as helicity injection
in our observations. 

The similar type of the dynamics in form of the parabolic trajectory has been reported 
in case of various types of spicules (type I and II), dynamic fibrils,
mottles (Hansteen et al. 2006; De Pontieu et al. 2007a,b, 2011).
These findings also support the acceleration of such small-scale 
jets by the shock triggered in the chromosphere due to velocity
pulses generated by the leakage of sub-photospheric p-modes. However,
such drivers need few km s$^{-1}$ velocity amplitude,
3-5 minutes temporal scales, as well as some concentrated 
field regions of few hundred Gauss to push such jets 
2-10 Mm above the solar limb.
Recently, Murawski \& Zaqarashvili (2010)
have also simulated the solar spicules by launching 
a velocity pulse above the solar photosphere.
However, the strong velocity pulse will be needed in the present case 
to trigger a jet material higher in the corona. 

Our observed dynamics supports that this jet may also be formed by a velocity
pulse that triggered in the chromosphere and steepens in form of a shock
in the transition region/corona. However, it may not be associated with
typical photospheric velocity pulses of the order of 
few km s$^{-1}$ associated with the leakage of p-modes 
that launches the small-scale jets as mentioned in many significant 
previous studies (Hansteen et al. 2006; De Pontieu et al. 2007a,b). In the present case, the steepened shock can be 
generated by a reconnection generated strong velocity pulse. Its front may rarefy 
the plasma density behind it, and the jet plasma could move up.
After reaching in the corona along a curved magnetic field lines, 
this jet falls back due to gravity and cool plasma
traced back its backward path as observed here.
In next section, we
discuss in details the numerical simulation of the observed jet in the solar
atmosphere triggerred by a strong velocity pulse.

\section{A numerical model}\label{SECT:NUM_MODEL}
%
%
%
Our model system is taken to be composed of
a gravitationally-stratified solar atmosphere 
which is described by
the ideal two-dimensional (2D) 
MHD equations:
\beqa
\label{eq:MHD_rho}
{{\partial \varrho}\over {\partial t}}+\nabla \cdot (\varrho{\bf V})=0\, ,
\\
\label{eq:MHD_V}
\varrho{{\partial {\bf V}}\over {\partial t}}+ \varrho\left ({\bf V}\cdot \nabla\right ){\bf V} =
-\nabla p+ \frac{1}{\mu}(\nabla\times{\bf B})\times{\bf B} +\varrho{\bf g}\, ,
\\
\label{eq:MHD_p}
{\partial p\over \partial t} + \nabla\cdot (p{\bf V}) = (1-\gamma)p \nabla \cdot {\bf V}\, ,
\\
\label{eq:MHD_B}
{{\partial {\bf B}}\over {\partial t}}= \nabla \times ({\bf V}\times{\bf B})\, , 
\hspace{3mm}
\nabla\cdot{\bf B} = 0\, .
\eeqa
Here ${\varrho}$ is mass density, ${\bf V}$ is flow velocity,
${\bf B}$ is the magnetic field, $p = \frac{k_{\rm B}}{m} \varrho T$ is gas pressure, $T$ is temperature,
$\gamma=5/3$ is the adiabatic index, ${\bf g}=(0,-g)$ is gravitational acceleration of
its value $g=274$ m s$^{-2}$,
$m$ is mean particle mass and $k_{\rm B}$ is the Boltzmann's constant.
\subsection {Equilibrium configuration}
%
%
%
We assume that at its equilibrium the solar atmosphere is still (${\bf V}_{\rm e}=0$) with a force-free magnetic field,
\begin{equation}\label{eq:B_e}
\label{eq:B}
(\nabla\times{\bf B}_{\rm e})\times{\bf B}_{\rm e} = 0\ , 
\end{equation}
such that it satisfies a current-free condition,
%
$\nabla \times \vec B_{\rm e}=0$, and it is specified by the magnetic flux function, $A$,
as
$$
\vec B_{\rm e}=\nabla \times (A\hat {\bf z})\, .
$$
%
Here the subscript $_{\rm e}$ corresponds to equilibrium quantities.
We set an arcade magnetic field by choosing
%
\begin{equation}
A(x,y) = B_{\rm 0}{\Lambda}_{\rm B}\cos{(x/{\Lambda}_{\rm B})} {\rm exp}[-(y-y_{\rm r})/{\Lambda}_{\rm B}]\, .
\end{equation}
%
%
%
%
Here, $B_{\rm 0}$ is the magnetic field at $y=y_{\rm r}$, and the magnetic scale-height is
\beq
{\Lambda}_{\rm B}=2L/\pi\, .
\eeq
%
We set and hold fixed $L=100$ Mm. 

As a result of Eq. (5) 
the pressure gradient is balanced by the gravity force,
\begin{equation}
\label{eq:p}
-\nabla p_{\rm e} + \varrho_{\rm e} {\bf g} = 0\, .
\end{equation}
%
%
%
With a use of the ideal gas law and the $y$-component of 
Eq. (8)
, we 
arrive at 
\beqa
\label{eq:pres}
p_{\rm e}(y)=p_{\rm 0}~{\rm exp}\left[ -\int_{y_{\rm r}}^{y}\frac{dy^{'}}{\Lambda (y^{'})} \right]\, ,\hspace{3mm}
\label{eq:eq_rho}
\varrho_{\rm e} (y)=\frac{p_{\rm e}(y)}{g \Lambda(y)}\, ,
\eeqa
where
\begin{equation}
\Lambda(y) = k_{\rm B} T_{\rm e}(y)/(mg)
\end{equation}
is the pressure scale-height, and $p_{\rm 0}$ denotes the gas 
pressure at the reference level that we choose in the solar corona at $y_{\rm r}=10$ Mm.

We adopt
an equilibrium temperature profile $T_{\rm e}(z)$ for the solar atmosphere
that is close to the VAL-C atmospheric model of Vernazza et al. (1981). 
Then with the use of Eq. (9)
 we obtain the corresponding gas pressure and mass density profiles.

%
%

%
%
\subsection{Perturbations}
%
We initially perturb
the above equilibrium impulsively by a Gaussian pulse in the
vertical component of
velocity, $V_{\rm y}$, 
viz.,
\beq\label{eq:perturb}
V_{\rm y}(x,y,t=0) = A_{\rm v} 
\exp\left[ 
-\frac{(x-x_{\rm 0})^2} {w_{\rm x}^2}
-\frac{(y-y_{\rm 0})^2} {w_{\rm y}^2} 
\right]\, .
\eeq
Here $A_{\rm v}$ is the amplitude of the pulse, $(x_{\rm 0},y_{\rm 0})$ is its initial position and
$w_{\rm x}$, $w_{\rm y}$ denote its widths along the $x$- and $y$-directions, respectively. 
We set and hold fixed $A_{\rm v}=160$ km s$^{-1}$, $x_{\rm 0}=-17.5$ Mm, $y_{\rm 0}=1.75$ Mm, 
$w_{\rm x}=10$ Mm, and $w_{\rm y}=0.1$ Mm. 
%
%
%
\section{Results of numerical simulations}
%
\begin{figure*}
\centering
\includegraphics[width=5.0cm,height=8.0cm, angle=0]{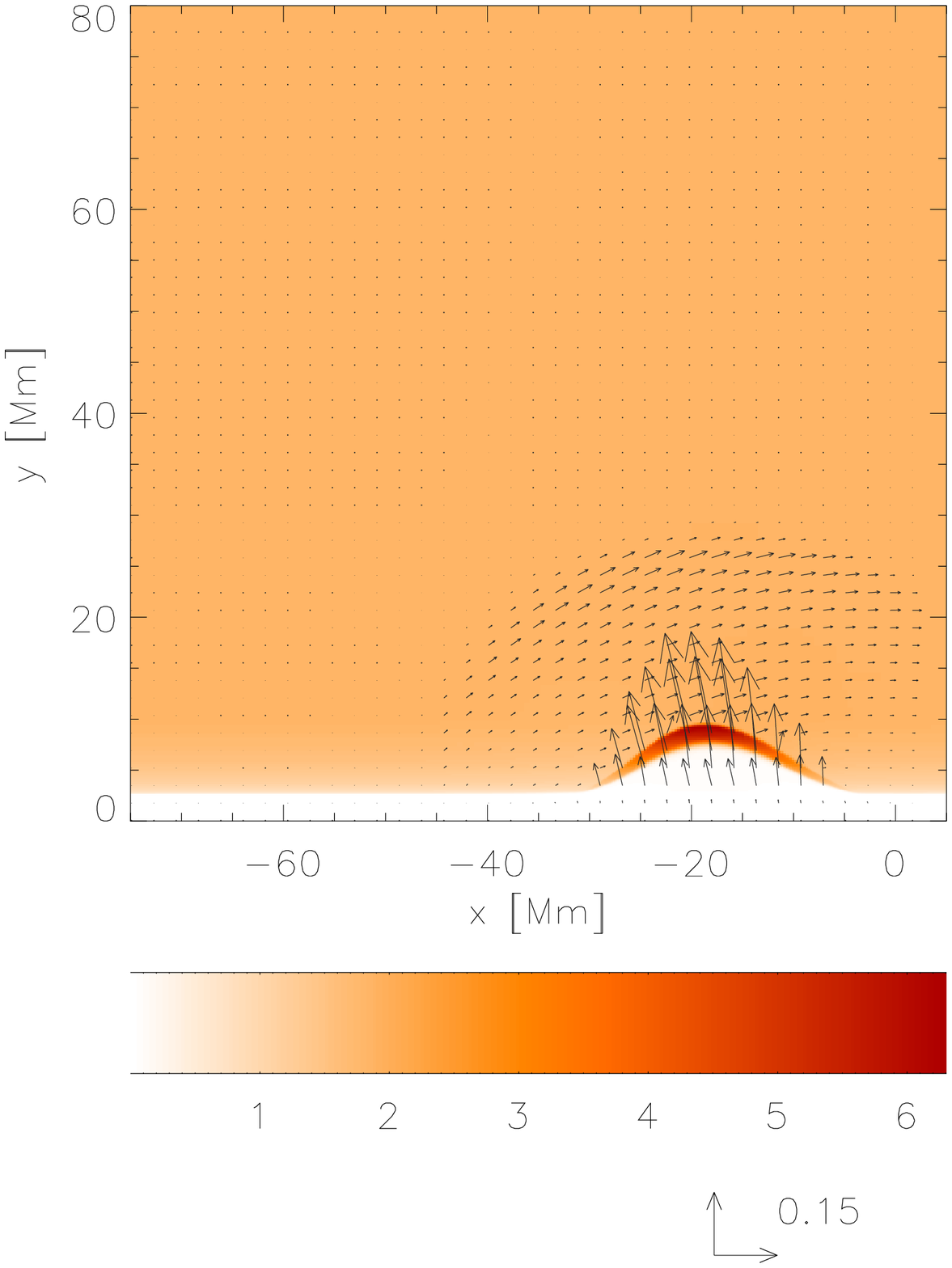}
\includegraphics[width=5.0cm,height=8.0cm, angle=0]{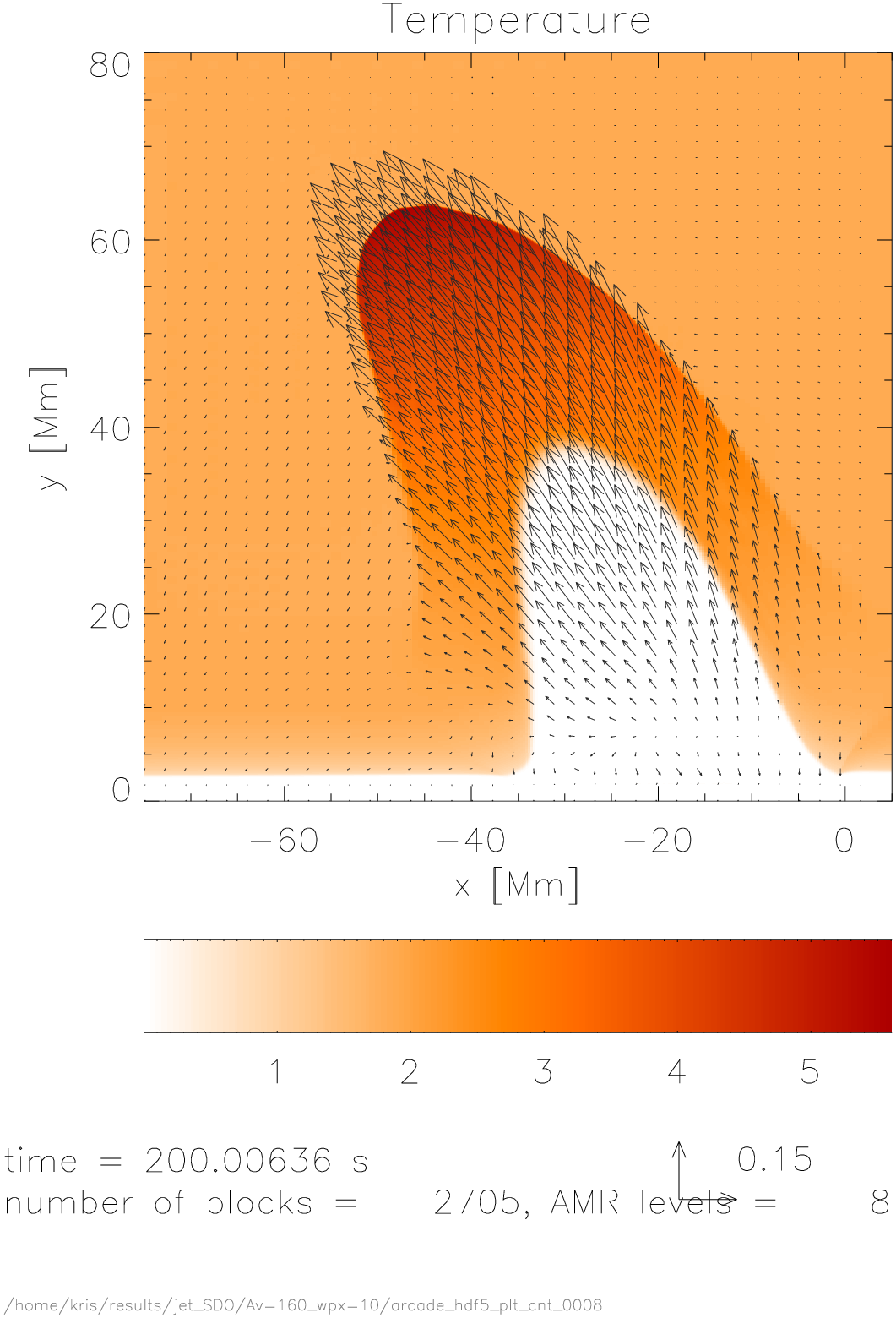}
\includegraphics[width=5.0cm,height=8.0cm, angle=0]{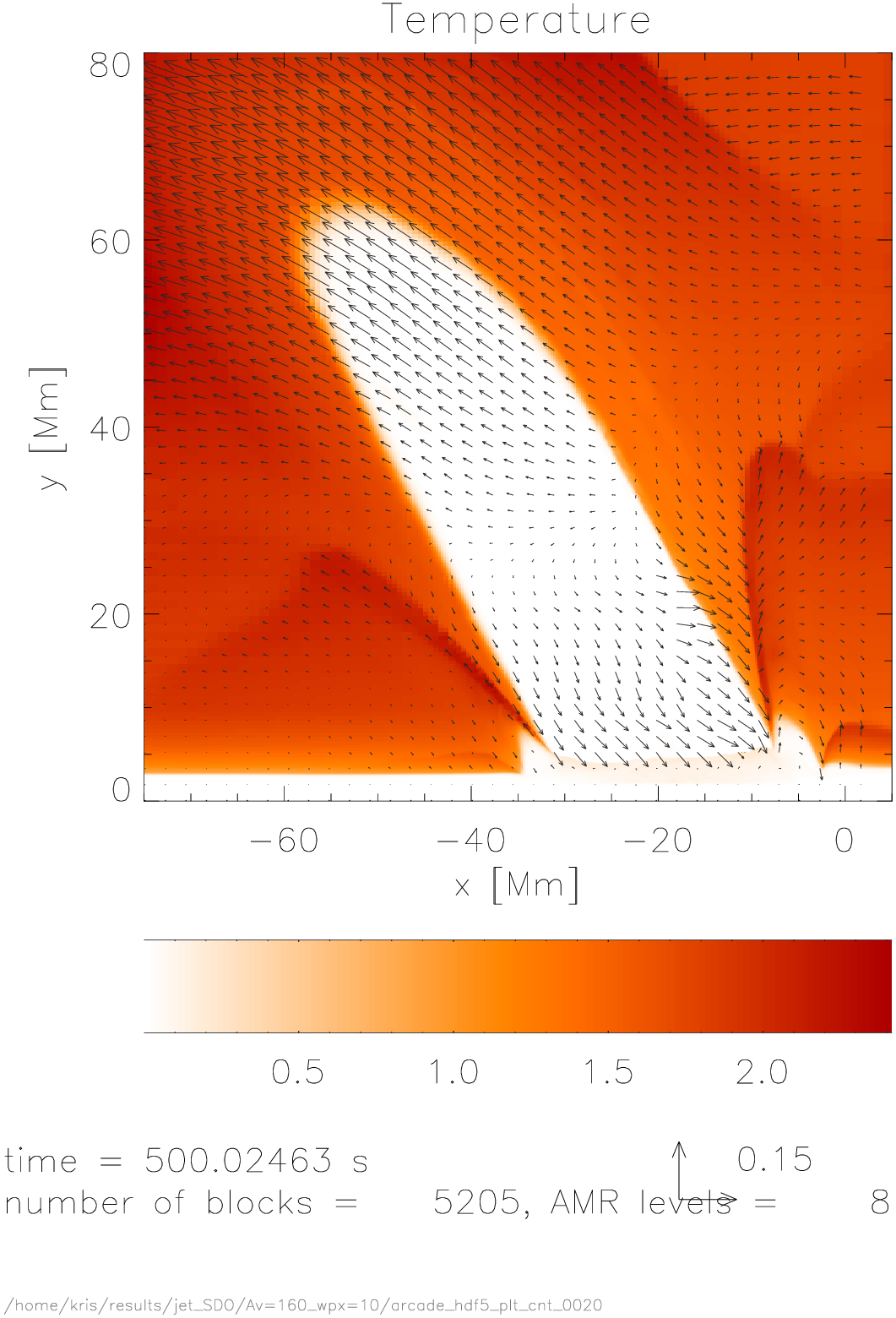}\\
\includegraphics[width=5.0cm,height=8.0cm, angle=0]{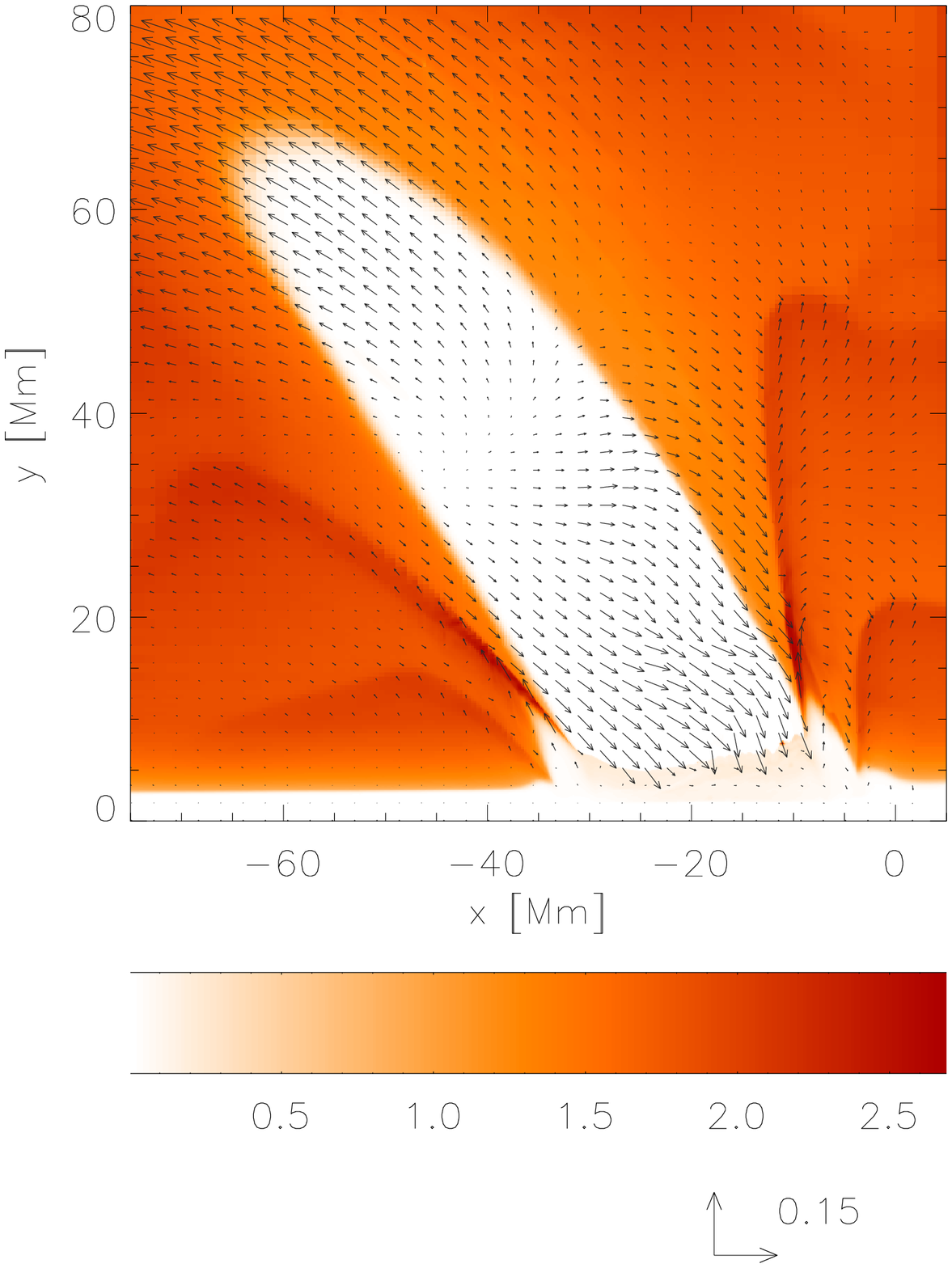}
\includegraphics[width=5.0cm,height=8.0cm, angle=0]{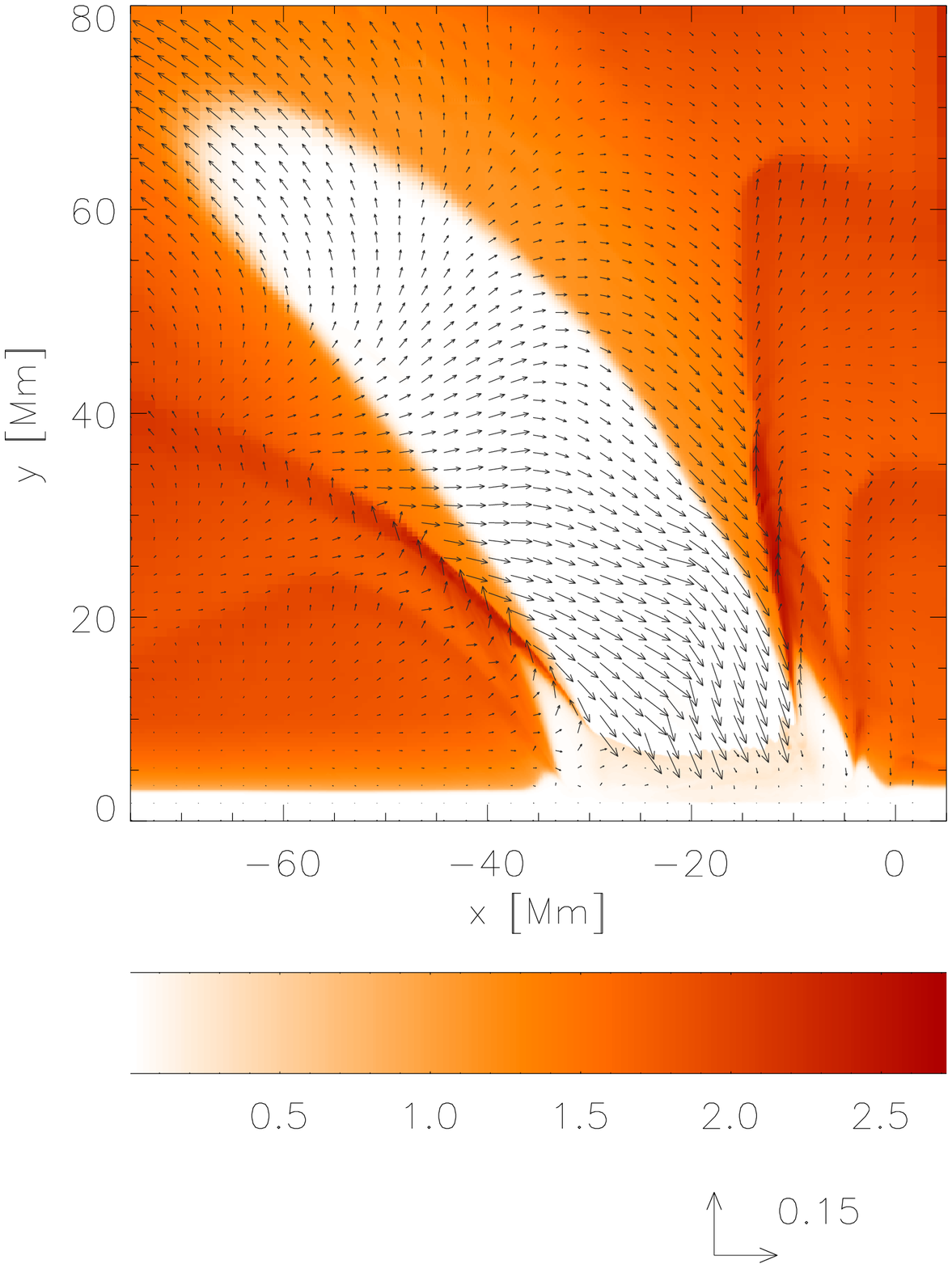}
\includegraphics[width=5.0cm,height=8.0cm, angle=0]{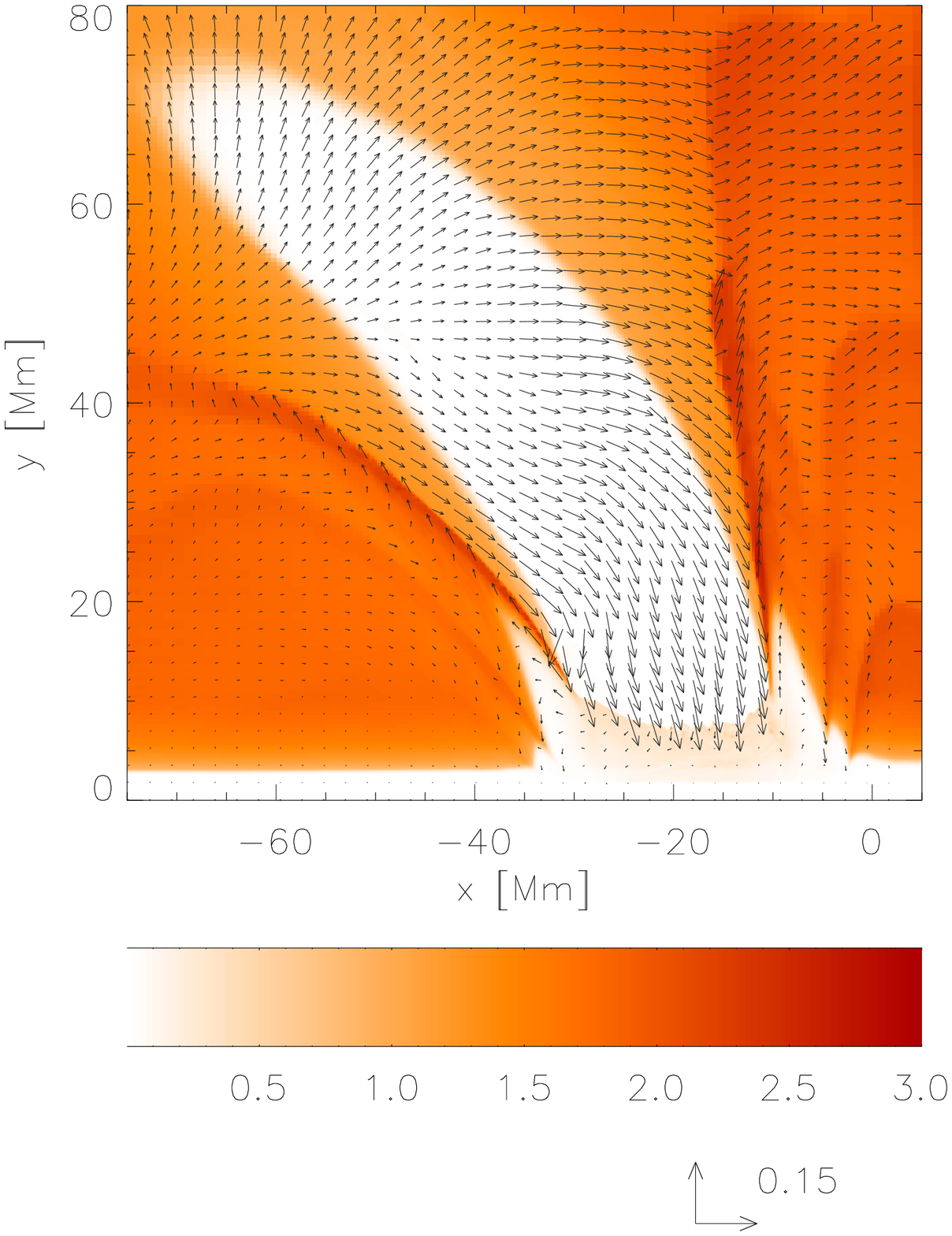}\\
\includegraphics[width=5.0cm,height=8.0cm, angle=0]{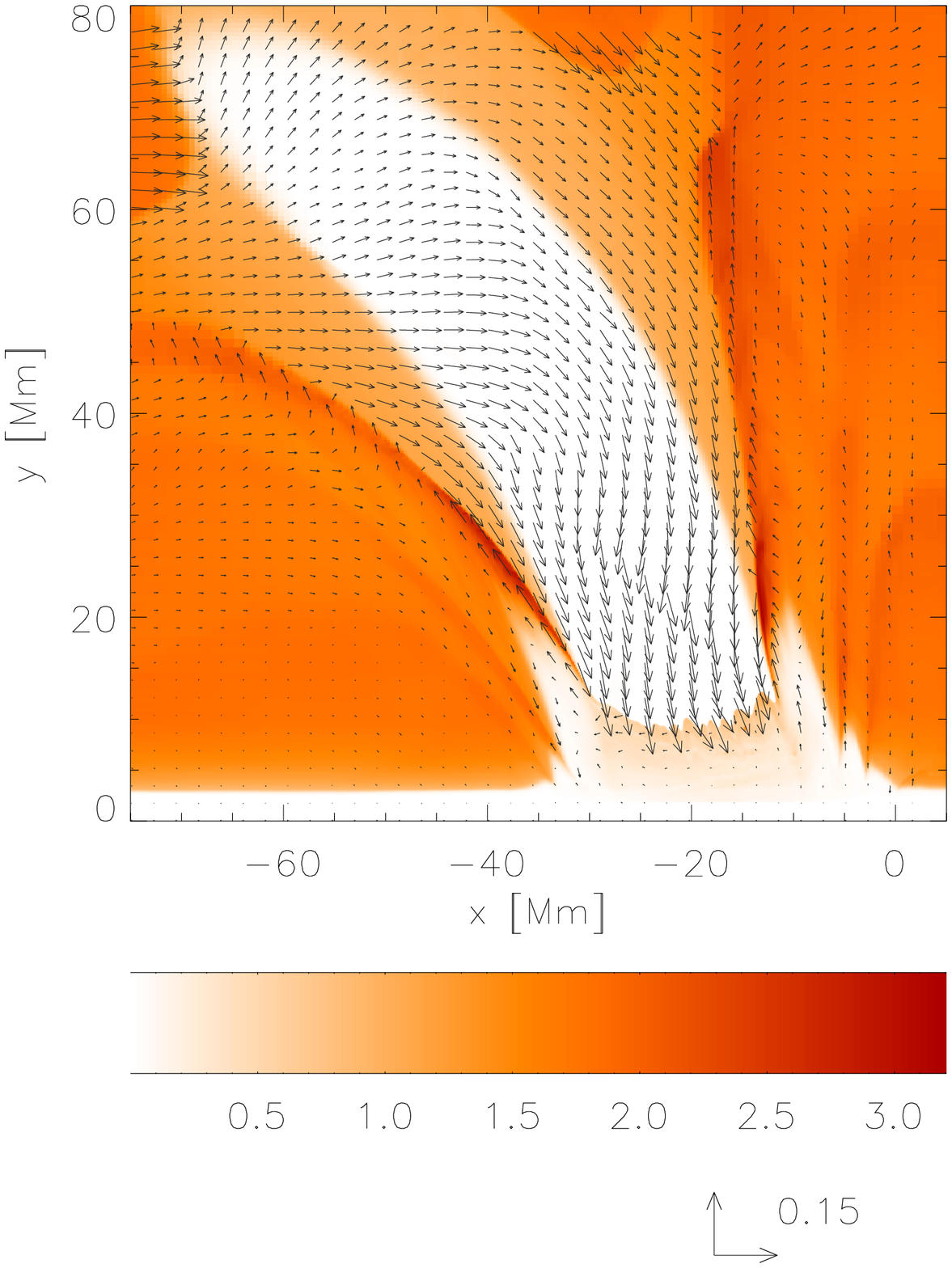}
\includegraphics[width=5.0cm,height=8.0cm, angle=0]{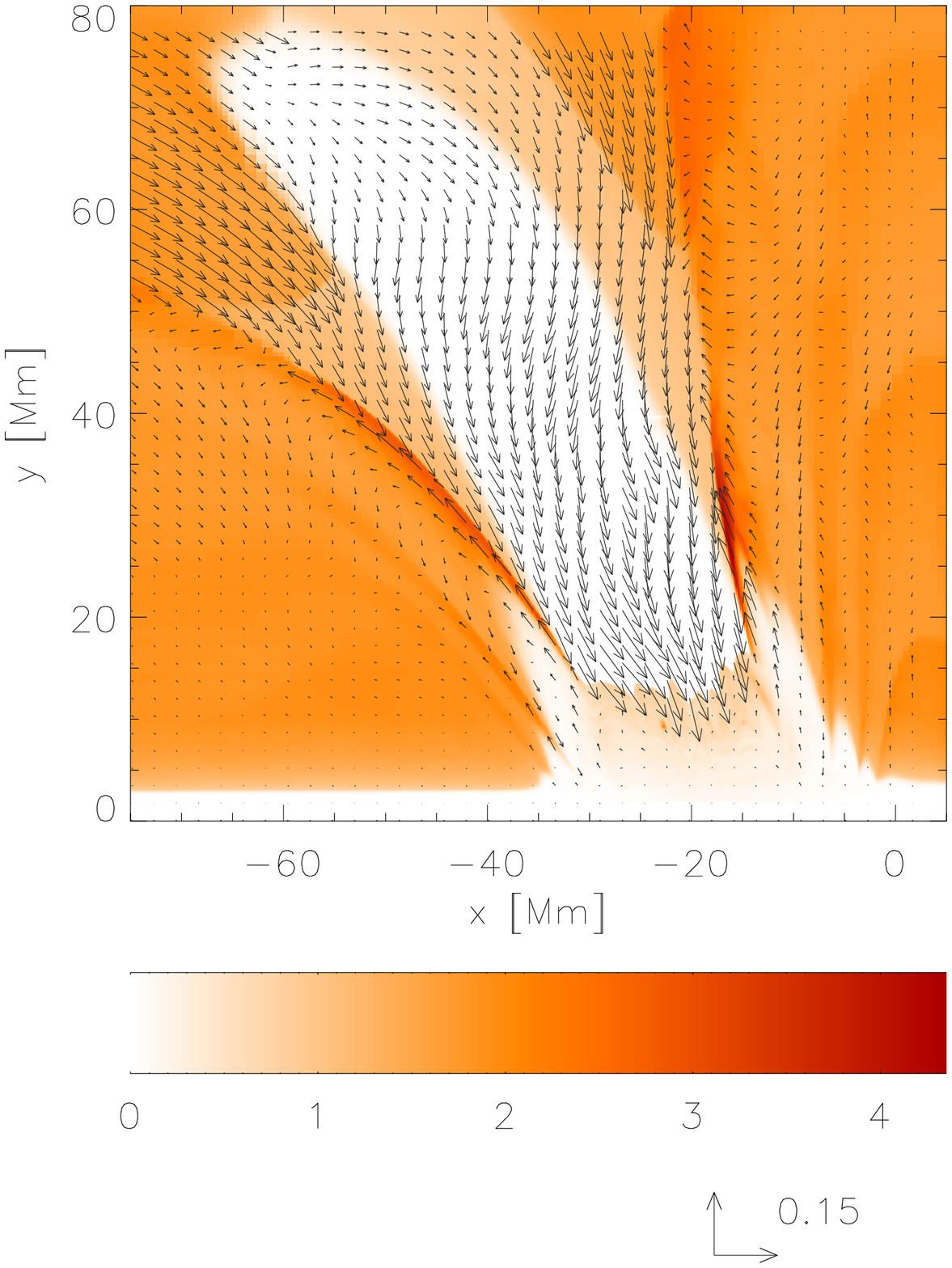}
\includegraphics[width=5.0cm,height=8.0cm, angle=0]{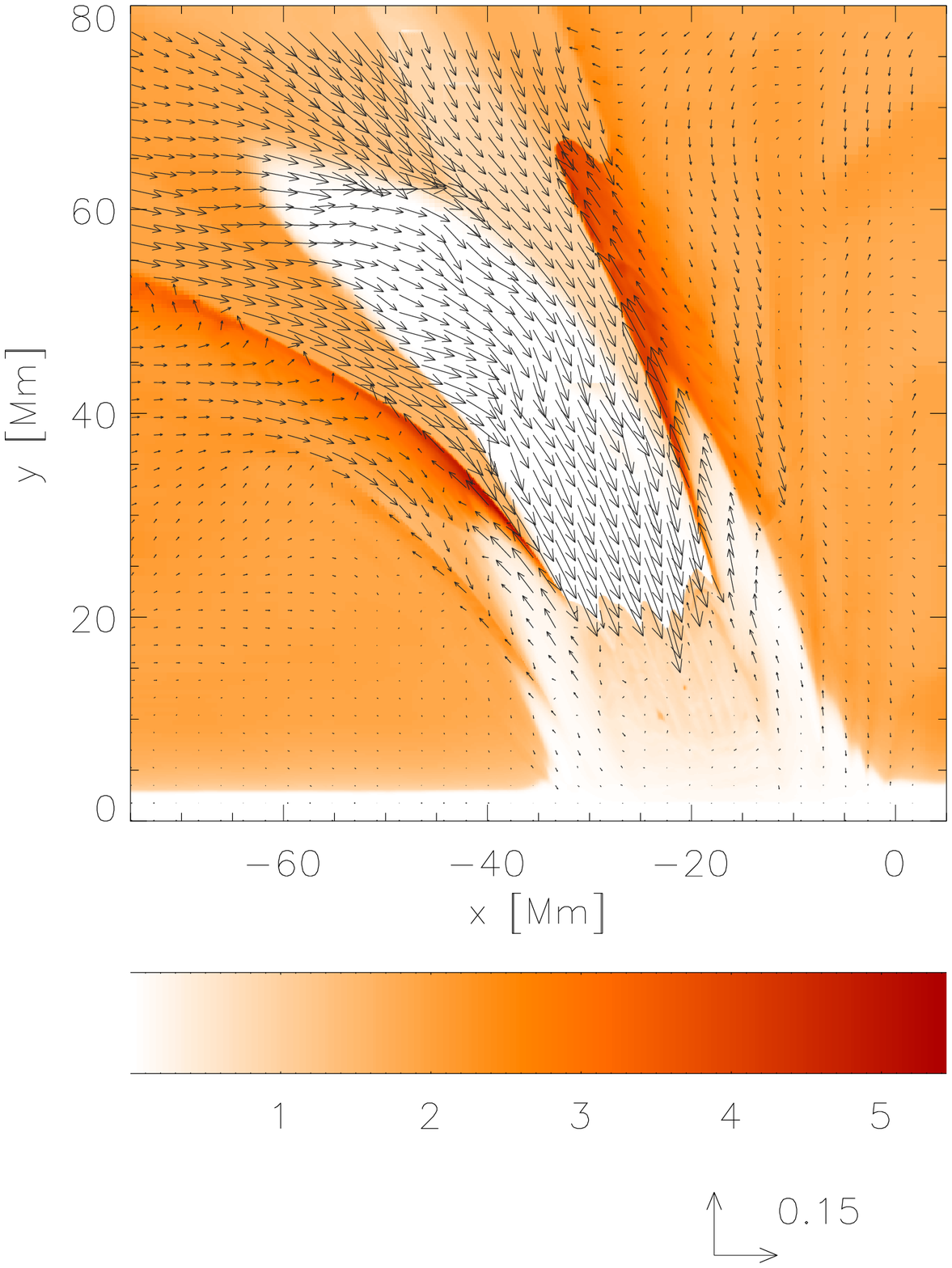}\\
\caption{\small 
Temperature (colour maps) and velocity (arrows) profiles at
$t=25$ s, $t=200$ s, $t=500$ s, 
$t=600$ s, $t=700$ s, $t=800$ s, 
$t=900$ s, $t=1000$ s, and $t=1200$ s
(from top-left to bottom-right). 
Temperature is drawn in units of $1$ MK. 
The arrow below each panel represents the length of the velocity vector, expressed in units of $150$ km s$^{-1}$. 
}
\label{fig:jet_prof}
\end{figure*}

Equations (1)-(4)
 are solved numerically using the code FLASH
(Lee \& Deane 2009). This code implements a second-order unsplit Godunov solver 
with various slope 
limiters and Riemann solvers, as well as adaptive mesh refinement (AMR).
We set the simulation box 
of $(-125,15)\, {\rm Mm} \times (0,110)\, {\rm Mm}$ along the $x$- and $y$-directions 
and impose fixed in time all plasma quantities at all four boundaries of the simulation region. 
In all our studies we use AMR grid with a minimum (maximum) level of 
refinement set to $4$ ($8$). The refinement strategy is based on 
controlling numerical errors in 
mass density, which results in an excellent resolution of steep spatial profiles and
greatly reduces numerical diffusion at these locations.

Figure 4 
displays the spatial profiles of plasma temperature (colour maps) and velocity (arrows) 
resulting from the initial velocity pulse of Eq. (11)
, which splits into counter-propagating parts. The part which propagates downwards 
becomes reflected from the dense plasma layers at the photospheric region. This reflected part lags behind the originally upward propagating 
signal which becomes a shock. As the plasma is initially pushed upwards the under-pressure results in the region below the initial pulse. This under-pressure sucks up 
comparatively cold chromospheric plasma which lags behind the shock front at higher temperature. 
As a result, the pressure gradient force works against gravity and forces 
the chromospheric material to penetrate into the solar corona in form of rarefaction wave. 
At $t=25$ s this shock reaches the altitude of $y\simeq 10$ Mm and the rarefaction wave is
located 1 Mm below the slow shock. 
%
The next snapshot (top middle panel) is drawn for $t=200$ s. At this time the slow shock reached the altitude of $y\simeq 63$ Mm 
while the chromospheric plasma is located below at $y\simeq 40$ Mm. 
At the next moment of time, $t=500$ s, (top right panel) the shock moved up 
and the chromospheric plasma blob exhibits its developed phase, reaching the level of $y=64$ Mm, 
which matches well with the observational data of Fig. 3
 (top right panel).
The cool jet slow down while propagating upwards. At $t=800$ s it arrived at the level of $y\simeq70$ Mm (Fig.3, middle-right
panel)and subsequently 
subsided 
as the plasma began to flow downward,
being attracted by gravity (bottom panels).
At the subsiding stage the fine structures developed as a 
result of interaction between falling down plasma 
and rising up secondary pulse. Similar features were evident
in the case of the numerical simulation of spicules (Murawski \& Zaqarashvili 2010). 

\section{Discussions and Conclusion}\label{SECT:DISS}
Our numerical simulations show a general scenario of rising and subsequently falling down cool plasma jet with a number of 
observational features like jet height and width, ballistic motion, approximate rising time-scale and average rising velocity that are being close to the observational data. However, some match between the numerical 
and observational data is approximate and rather qualitative, e.g., the rising time-scale in the simulation is $\sim$800 s while in the real observations it is $\sim$720 s. This mismatch may obviously result from some of the simplified profiles of the model parameters in our developed numerical model, e.g., the magnetic field configuration at a place at which the initial pulse 
was launched, and so on. Moreover, the real jet was excited in more complex plasma and magnetic field conditions at the polar coronal hole of the Sun. We might not model exactly realistically the way in which the jet was excited in the real Sun. In fact, the exciter could work for some time, it could be located at a different place and it could have a different size, speed and distribution. As such data is not provided by the observations we decided to trigger the jet by a localized pulse in a plasma velocity, that is launched below the transition region. By this way we managed to excite the jet which mimics on average the properties of the observed cool plasma jet. The small mismatches like the exact height and length of the jet as well as its descending time can be reduced by tuning of free parameters in the numerical model. In particular, the jet length would be smaller for a smaller amplitude of the initial pulse. We did not expect to fit perfectly our numerical data to the observational findings but instead our intention was to show a qualitative and to some extent quantitative agreement between the numerical and observational approaches.

It is noteworthy that recently Culhane et al. (2007) have found the first observational evidence of unique EUV polar jets which could not escape from the Sun as like the jets previously observed with SXT in Yohkoh era. They conject that such accelerated polar jet plasma is being heated upto coronal temperature with its rise, and then fall back to lower atmosphere after its cooling. They observed jet velocities in the range of 150 km s$^{-1}$- 360 km s$^{-1}$ and reported their formation most probably due to the magnetic reconnection between emerging bipoles with the large-scale open field lines of the polar coronal hole. Ko et al. (2005) have also found cool and hot plasma components in a limb jet, as well as its falling back scenario in the lower atmosphere. However, all these previously observed jets were found to be consistent with the impulsive transient heating (e.g., as driven by direct reconnection processes), and thereafter cooling and draining along the same path. We also observe the jet motion that moves up and thereafter falls along the same path in the lower atmosphere. However, the jet is made by the cool plasma maintained at a temperature of $\sim$10$^{5}$ K, which is sensitive to the SDO/AIA 304 \AA\ filter representing upper chromospheric/TR plasma. However, the jet plasma column is not evident at typical coronal temperature of 1.0 MK in the SDO/AIA 171 \AA\ . Therefore, the jet consists of a cool plasma even after reaching at the coronal heights, and does not subject to any heating and cooling events as previously observed by Culhane et al. (2007); Ko et al. (2005). The primary mechanism for the origin of such jet may be the magnetic reconnection between the emerging bipolar magnetic fields and pre-existing open field lines in the polar coronal hole (Culhane et al. 2007). However, comparatively slow average rising speed (110 km s$^{-1}$), steady ballistic motion, the absence of impulsive heating at the base and no gradual enhancement in plasma temperature etc., indicate the role of magnetic reconnection as an indirect mechanism. The jet is seem to be driven by a large amplitude velocity pulse that may be excited non-linearly by recurrent magnetic reconnection at the base of the polar coronal hole. This pulse steepens into a slow shock front which moves at
a higher temperature wave front. However, the density and typical emissions are low in polar coronal holes. Therefore, we could not observe high temperature bright front of the jet. Later, the motion of the slow shock, under stable conditions, causes the propagative rarefaction wave which direct the cool and denser chromospheric/TR plasma upward. This steadily moving plasma is clearly visible in the observations as well as in the simulation. After reaching at the height of $\sim$75 Mm, the plasma begins to subside. The gravitational free fall plays the role for the falling of jet material. 
During the subsiding phase of the jet, in the numerical simulation, the another pulse is evident at the base that causes the low rate of subsided plasma compared to the observations. However, the realistic solar atmosphere is rather complex, and there is a possibility that the huge downfalling material causes the vanishing of the another pulses coming from the lower atmosphere. This is the reason why a pulse driven single jet became evident above the limb rather than multiple jets at the same place.  

In conclusion we suggest that the initial velocity pulse launched below the transition region is able to trigger a hot plasma shock which is followed by a cool plasma jet. This cold plasma jet approximately resembles many of the features which are found in the presented SDO/AIA observational data. We report first time on the observations of a pulse driven plasma jet in the polar region and provide theoretical explanation of this phenomenon on the basis of numerical 
simulations we performed. However, further multiwavelength observations should be performed by high-resolution space borne (e.g., SDO, Hinode, STEREO) and complementary ground based observations to shed new light on this kind of unique jets that rise and fall back in the solar atmosphere. This will also impose a rigid constraint on the stringent simulations of such kind of jets in the model solar atmosphere.

\section{Acknowledgment}
We thank the reviewer for his/her valuable suggestions that improved the manuscript considerably.
We acknowledge the use of the SDO/AIA observations for this study. The data is provided curtesy of NASA/SDO, LMSAL, and the AIA, EVE, and HMI science teams. The FLASH code has been developed by the DOE-supported ASC/Alliance
Center for Astrophysical Thermonuclear Flashes at the University of
Chicago.
AKS thanks Shobhna Srivastava for patient encouragements during the work.
%
%

\end{document}